# Capturing electron-driven chiral dynamics in UV-excited molecules

Vincent Wanie[1*], Etienne Bloch[2], Erik P. Månsson[1], Lorenzo Colaizzi[1,3], Sergey Ryabchuk[3,4], Krishna Saraswathula[1,3], Andres F. Ordonez[5], David Ayuso[5,6], Olga Smirnova[6,7], Andrea Trabattoni[1,8], Valérie Blanchet[2], Nadia Ben Amor[9], Marie-Catherine Heitz[9], Yann Mairesse[2], Bernard Pons[2*], Francesca Calegari[1, 3, 4*]

[1]Center for Free-Electron Laser Science, Deutsches Elektronen-Synchrotron DESY, Notkestr. 85, 22607 Hamburg, Germany
[2]Université de Bordeaux - CNRS - CEA, CELIA, UMR5107, F-33405 Talence, France
[3]Physics Department, Universität Hamburg, Luruper Chaussee 149, 22761 Hamburg, Germany
[4]The Hamburg Centre for Ultrafast Imaging, Universität Hamburg, Luruper Chaussee 149, 22761 Hamburg, Germany
[5]Department of Physics, Imperial College London, SW7 2AZ London, UK
[6]Max-Born-Institut, Max-Born-Str. 2A, 12489 Berlin, Germany
[7]Technische Universität Berlin, Straße des 17. Juni 135,10623 Berlin, Germany
[8]Institute of Quantum Optics, Leibniz Universität Hannover, Welfengarten 1, 30167 Hannover, Germany
[9]CNRS, UPS, LCPQ (Laboratoire de Chimie et Physique Quantiques), FeRMI, 118 Route Narbonne, F-31062 Toulouse, France
*Email: vincent.wanie@desy.de; bernard.pons@u-bordeaux.fr; francesca.calegari@desy.de

**Molecular chirality is a key design property for many technologies including bioresponsive imaging[1], circularly polarized light detection[2] and emission[3], molecular motors[4,5] and switches[6,7]. Imaging and manipulating the primary steps of transient chirality is therefore central for controlling numerous physical, chemical and biological properties that arise from chiral molecules in response to external stimuli. So far, the manifestation of electron-driven chiral dynamics in neutral molecules has not been demonstrated at their intrinsic timescale. Here, we use time-resolved photoelectron circular dichroism (TR-PECD)[8–12] with an unprecedented instrument response function of 2.9 fs to image the dynamics of coherent electronic motion activated by prompt UV-excitation in neutral chiral molecules, disclosing its impact on the molecular chiral response. We find that electronic beatings between Rydberg states lead to periodic modulations of the chiroptical response on the few-femtosecond timescale, showing a sign inversion in less than 10 fs. Calculations including both the molecular UV-excitation and subsequent photoionization confirm this interpretation and provide further evidence that the combination of the resulting photoinduced chiral current with a circularly polarized probe pulse realizes an enantio-selective filter of molecular orientations upon photoionization, opening up a route towards enantio-selective charge-directed reactivity[13].**

Chirality is a peculiar property that characterizes the majority of biochemical systems: a chiral molecule exists in two geometrical configurations that are non-superimposable mirror images of each other – defined as (*R)* and (*S*) enantiomers – exhibiting different physical and chemical properties when interacting with another chiral entity. This chiral recognition is central to many fields of applied sciences[14], including enantio-selective catalysis[15], drug engineering and biophysics[7]. Capturing the primary steps of chiral recognition and the mechanisms dictating the outcomes of a chiral interaction would thus have a significant impact in various fields dealing with chiral properties of matter. At the ultrafast electronic timescale, the opportunity to steer electrons responsible for chemical activity notably promises a way to control the outcome of enantio-sensitive phenomena through the concept of charge-directed reactivity[16–18]. Electron currents in photoexcited chiral molecules have indeed been identified as precursors for driving enantio-sensitive molecular orientation[13]. Nonetheless, the ability to track and manipulate electron-driven chiral interactions in neutral molecules is still pending.

In this context, the temporal resolution provided by attosecond technologies developed in the past twenty-two years gives access to some of the fastest electronic dynamics of matter on their natural timescale. Seminal pump-probe experiments using attosecond light pulses have revealed valence electron dynamics in atoms[19], autoionization dynamics in molecules[20], photoionization delays in solids[21,22] as well as electron-driven charge migration in *ionized* biomolecules[18,23]. In all these cases, the intrinsically high photon energy of the attosecond light sources inevitably leads to ionization of the target, restricting the measurements to ultrafast dynamics of cationic states.

Instead, investigating the light-induced electron dynamics of biochemically-relevant chiral molecules in their *neutral* states with high temporal resolution requires new experimental approaches, and important considerations must be taken into account. First, the pump pulse must have well-defined characteristics: (i) a photon energy below the ionization threshold, (ii) a broadband energy spectrum to trigger coherent electron motion among multiple electronic states and (iii) a time duration that provides a prompt excitation before any nuclear motion can take place, together with sufficient temporal resolution. Because of the low ionization potential of most molecular systems, laser pulses with such characteristics are confined to the spectral region covering the ultraviolet (UV) and vacuum-UV ranges, which also avoids triggering intricate high-order, strong-field multiphoton driven processes that do not exist in nature[24,25]. All these characteristics would provide the tools to harness temporal resolutions previously unattained in pump-probe spectroscopic techniques that are highly sensitive to chirality, such as time-resolved photoelectron circular dichroism (TR-PECD)[8–12].

Here, we meet all the above requirements by using - for the first time - ultrashort UV pump pulses[26] in combination with circularly polarized near-infrared (NIR) probe pulses, to study coherent electronic dynamics in chiral neutral molecules with unprecedented temporal resolution. We apply the chiroptical method of TR-PECD to investigate electron-driven chiral interactions in neutral methyl-lactate (ML) $C_4H_8O_3$ – a derivative of lactate, which has regained substantial interest due to its recently uncovered metabolic functions[27]. Fig. 1(a-b) shows an overview of the experimental approach. First, a linearly polarized UV pulse promptly launches a coherent electronic wavepacket just below the ionization threshold in the bio-relevant molecule via a two-photon transition. Then, a time-delayed circularly polarized NIR probe triggers ionization from the transient wavepacket, providing an exceptional instrument response function of 2.90 ± 0.06 fs. For each pump-probe delay $t$, the 2D-projected photoelectron angular distributions (PAD) $S^{(h)}(\epsilon,\theta,t)$ are collected with a velocity map imaging spectrometer (VMIS), for both left ($h = +1$) and right ($h = -1$) circular polarizations of the probe pulse. $\epsilon$ and $\theta$ stand for the kinetic energy and direction of ejection of the photoelectron in the (x,z) VMIS detection plan, respectively. The chiroptical response is characterized by a PECD image defined as the normalized difference $PECD(\epsilon,\theta,t) = 2\frac{S^{(+1)}(\epsilon,\theta,t)-S^{(-1)}(\epsilon,\theta,t)}{S^{(+1)}(\epsilon,\theta,t)+S^{(-1)}(\epsilon,\theta,t)}$, subsequently fitted using a pBasex inversion algorithm[8]. Snapshots of the measured $PECD(\epsilon,\theta,t)$ are presented in Fig. 1(c), discriminating between low ($\epsilon \leq 100$ meV) and high ($\epsilon >100$ meV) energy electrons. Low energy electrons are preferentially emitted in the $\theta = 180°$ backward hemisphere at $t = 5$ fs, and preferentially ejected forward at $t = 11$ fs. Their main direction of ejection reverses again at $t = 17$ fs. Higher energy electrons are more likely emitted forward for $t \geq 11$ fs, with a magnitude that depends on $t$.

The $PECD(\epsilon,\theta,t)$ images provide quantitative fingerprints of an ultrafast dynamics taking place on the few-femtosecond timescale. In order to further characterize the temporal evolution of the observed dynamics, we decompose the PAD images in series of Legendre polynomials, $S^{(h)}(\epsilon,\theta,t) = \sum_{n=0}^{6} b_n^{(h)}(\epsilon,t)P_n(\cos\theta)$, and calculate the multiphoton PECD[28], defined as the normalized difference of electrons emitted in the forward and backward hemispheres for $h = +1$, as $MP-PECD(\epsilon,t) = 2\beta_1^{(+1)}(\epsilon,t) - \frac{1}{2}\beta_3^{(+1)}(\epsilon,t) + \frac{1}{4}\beta_5^{(+1)}(\epsilon,t)$ where $\beta_n^{(+1)}(\epsilon,t) = \frac{b_n^{(+1)}(\epsilon,t)}{b_0^{(+1)}(\epsilon,t)}$ (see Methods). $\beta_1^{(+1)}(\epsilon,t)$ refers to the isotropic part of the asymmetry in each hemisphere, whereas $\beta_3^{(+1)}(\epsilon,t)$ encodes anisotropic features due to primary excitation[8], leading to the angular shaping of the PECD illustrated in Fig. 1(c) – $\beta_5^{(+1)}(\epsilon, t)$ has been found negligible in our measurements. Figure 2(a) shows $MP-PECD(\epsilon, t)$ while its $\beta_1^{(+1)}(\epsilon, t)$ component is displayed in (b). The results are shown for (*S*)-ML and a mirroring symmetric measurement in (*R*)-ML clearly confirms the chiral character of the Rydberg-induced dynamics, with minor discrepancies due to slightly lower enantiopurity and statistics (see Fig. S4 of the Supplementary Information, SI). We observe that $\beta_1^{(+1)}(\epsilon, t)$ closely matches the MP-PECD behavior of Fig. 2(a), indicating that the subtle anisotropic effects included in $\beta_3^{(+1)}(\epsilon, t)$ play a minor role. The multiphoton MP-PECD can be partitioned into three kinetic energy ranges, as identified in Fig. 2(a). Between 25 and 100 meV, the photoelectron emission asymmetry strikingly reverses in $\sim 7$ fs (see Fig. 2(c)). A clear modulation of the asymmetry remains over few tens of fs, which is also observed at higher $\epsilon$ between 100 and 300 meV (Fig. 2(d)) and 300 and 720 meV (Fig. 2(e)). These modulations are also visible in the time-resolved photoelectron yield $\beta_0^{(+1)}(\epsilon, t)$, albeit their contrast is considerably weaker (see Section 1.2 of the SI). This highlights the capabilities of TR-PECD, which relies on differential measurements, over conventional photoelectron spectroscopy. In the following, we aim at assigning the origin of the fast temporal modulation of the asymmetry, which could involve electronic and/or nuclear degrees of freedom.

We modeled the experiment including both the two-photon UV-excitation and the NIR photoionization steps as sequential perturbative processes, within the frozen-nuclei approximation. A detailed description of the theoretical model is provided in the Methods. The electronic spectrum of ML and the two-photon excitation amplitudes are obtained via large-scale time-dependent density functional theory[29]. Ionization from the excited states is described using the continuum multiple scattering X$\alpha$ approach[30,31].

We present the results of our calculations in Fig. 3. The pump pulse populates excited states mainly stemming from excitation of the highest occupied molecular orbital (HOMO) of the ML ground state (see section 2.1 of the SI). Panel (a) shows the two-photon excitation cross sections associated with almost pure HOMO excitation to Rydberg states. Subsequent photoionization by the probe pulse leads to the emission of photoelectrons with kinetic energies $\epsilon = 250$ and 500 meV, respectively. These $\epsilon$ values are representative of the second and third energy ranges discriminated in the experimental data, respectively – the case of low-energy photoelectron dynamics ($\epsilon = 50$ meV) is discussed in Section 2.2.3 of the SI. Including the HOMO excited states of Fig. 3(a) in the dynamical calculations yields the time-resolved MP-PECD displayed in panels (b) and (d). The calculations are started at $t = 10$ fs to ensure no temporal overlap between the pump and probe pulses. The computed asymmetry presents clear modulations as a function of the pump-probe delay. The power spectra of the MP-PECD signals, obtained by Fourier analysis, are compared to their experimental counterparts in panels (c) and (e). An excellent agreement is found at $\epsilon$ = 250 meV where the oscillatory pattern of the MP-PECD is traced back to the pump-induced coherent superposition of 3d and 4p Rydberg states respectively located at $E_{3d} = 8.834$ eV and $E_{4p} = 9.120$ eV in Fig. 3(a). This coherent superposition leads to quantum beatings with ~ 15 fs period, associated with the ~300 meV energy difference between the states, which survive long after the pump pulse vanishes. We note that the most stable geometries of methyl-lactate do not possess any vibrational mode in the vicinity of 2200 cm$^{-1}$ (~ 15 fs)[32]. Similarly, the coherent superposition of 4p and 4d,f Rydberg states results in the oscillatory feature of the MP-PECD signal measured at $\epsilon$ = 500 meV. A small mismatch of ~60 meV is observed between the experimental and theoretical power spectra in Fig. 3(e). This mismatch is on the order of the error made in quantum chemistry computations of excited state energies. Overall, Fig. 3(b) and 3(c) unmistakably demonstrate that the electronic coherence of the intermediate Rydberg states, as identified in Fig. 3(a), modulates the molecular chiroptical response.

In our fixed-nuclei description, the electron coherences leading to oscillatory MP-PECD do not vanish and even lead to an overestimation of the MP-PECD amplitude at all delays $t$. On the contrary, the oscillations observed in the experimental MP-PECD (Fig. 2(b-d)) become damped after about 40 fs. The time it takes for decoherence to occur in photoexcited[24,33] and photoionized[34–41] molecules is currently the topic of extensive investigations. Electronic wavepackets are subject to three main decoherence sources: (i) the decrease of the overlap between nuclear wavepackets evolving across different electronic states, (ii) the dephasing of the different wavepacket components, and (iii) the relaxation of electronic state populations induced by non-adiabatic couplings[36]. Chiral molecules provide an opportunity to increase the sensitivity of such studies due to the differential nature of PECD measurements.

In order to interpret the slow decoherence observed in our experiment, we investigate its dependence on the nuclear degrees of freedom. Describing the coupled electron and nuclear dynamics in an energy range where tens of electronic states lie is beyond state-of-the-art theoretical approaches. Therefore, we alternatively performed ab initio molecular dynamics calculations on the ground state of cationic ML to which all the HOMO Rydberg states involved in the pump-probe dynamics correlate upon ionization. Nuclear dynamics on the cationic and high-lying Rydberg surfaces are expected to be similar since the outer Rydberg electron does not significantly penetrate the molecular structure[42]. As detailed in Methods and Section 2.3 of the SI, two main classes of trajectories showed up, converging towards two main isomeric forms of the ML cation (see Fig. S10). Within each class of trajectories, the Rydberg state energies of neutral ML were found to remain approximately parallel to each other and to the ML cation along the reaction path (see Figure S11 of the SI). This favors the overlap of the nuclear wavepackets associated with different electronic Rydberg states over an extended time duration and thus minimizes the role of decoherence mechanisms (i) and (ii). This also strengthens the frozen-nuclei assumption in the description of electronic quantum beatings which are dictated by electronic energy differences and should therefore remain basically the same from $t = 0$ fs onwards. We assign the source of decoherence in the present investigation to non-adiabatic dynamics, not only between the states populated by the pump but also with the lower-lying states reached by internal conversion soon after the prompt excitation. This explains the decreasing amplitude of MP-PECD oscillations in Fig. 2(c-e) that is also compatible with the ~ 40 fs lifetime encoded in the time-dependent photoelectron yield (Fig. S4 of the SI).

We have seen that the oscillations of the chiroptical response for $\epsilon$ = 250 meV results from the coherent superposition of two states, the 3d and 4p Rydberg states. We now investigate in more detail the role of these states in the chiroptical response. For a single molecular orientation $\widehat{R}$, the excited electron wavepacket reads, at time $t$ after the pump pulse vanishes, $\Phi(\widehat{R}, r, t) = \sum_{j=3d,4p} A_j(\widehat{R}) \Psi_j(r) \exp(-iE_j t/\hbar)$ where $A_i(\widehat{R})$ are the real two-photon transition amplitudes associated with the excited states $\Psi_j(r)$. The associated electron density can be partitioned as

$$\rho(\widehat{R}, r, t) = \rho_{incoh}(\widehat{R}, r) + \rho_{cross}(\widehat{R}, r) \cos[(E_{4p} - E_{3d})t/\hbar] \qquad (1)$$

where $\rho_{incoh}(\widehat{\boldsymbol{R}}, \boldsymbol{r}) = A_{3d}^2(\widehat{\boldsymbol{R}})\Psi_{3d}^2(\boldsymbol{r}) + A_{4p}^2(\widehat{\boldsymbol{R}})\Psi_{4p}^2(\boldsymbol{r})$ and $\rho_{cross}(\widehat{\boldsymbol{R}}, \boldsymbol{r}) = 2A_{3d}(\widehat{\boldsymbol{R}})A_{4p}(\widehat{\boldsymbol{R}})\Psi_{3d}(\boldsymbol{r})\Psi_{4p}(\boldsymbol{r})$. Figure 4(a) shows, for one selected orientation $\widehat{\boldsymbol{R}}$, the coherent part $\rho(\widehat{\boldsymbol{R}}, \boldsymbol{r}, t) - \rho_{incoh}(\widehat{\boldsymbol{R}}, \boldsymbol{r})$ of the electron density, oscillating back-and-forth along the molecular structure with a period $T = 2\pi\hbar/(E_{4p} - E_{3d})$ of 14.4 fs. Ionization of the 3d and 4p state superposition leads, after averaging over the orientations $\widehat{\boldsymbol{R}}$, to the total photoelectron yield which can be decomposed similarly to (1):

$$b_0^{(\pm 1)}(\epsilon, t) = b_{0_{incoh}}^{(\pm 1)}(\epsilon) + b_{0_{cross}}^{(\pm 1)}(\epsilon)\cos\left[(E_{4p} - E_{3d})t/\hbar\right]. \tag{2}$$

The computed yield is presented in Fig. 4(b) for $\epsilon = 250$ meV, showing how the coherent state superposition leading to $b_{0_{cross}}^{(\pm 1)}(\epsilon)\cos\left[(E_{4p} - E_{3d})t/\hbar\right]$ modulates the incoherent sum $b_{0_{incoh}}^{(\pm 1)}(\epsilon)$ of individual cross sections. The unnormalized MP-PECD can in turn be written as:

$$MP\text{-}PECD(\epsilon, t) = MP\text{-}PECD_{incoh}(\epsilon) + MP\text{-}PECD_{cross}(\epsilon)\cos\left[\frac{(E_{4p} - E_{3d})t}{\hbar} - \Delta\phi\right] \tag{3}$$

where the additional phase $\Delta\phi$ arises from the interference of the state-selective continuum partial wave amplitudes building the asymmetry of the photoelectron yield (see SI). As usual, this interference is washed out at the level of the total photoelectron yield[30,43]. The temporal evolution of the unnormalized two-state MP-PECD is shown in Fig. 4(c), from which we extract the time delay $\Delta t = 1.8$ fs associated with $\Delta\phi = 0.79$ rad. The MP-PECD reverses sign within one period of the oscillation since the asymmetries of single 3d- and 4p-mediated pathways, contributing to the incoherent MP-PECD (dashed red line in Fig. 4(c)) around which the coherent part oscillates, verify $|MP\text{-}PECD_{cross}(\epsilon)| > |MP\text{-}PECD_{incoh}(\epsilon)|$. A similar behavior is observed at lower kinetic energy in the measurement reported in Fig. 2(c). Importantly, the MP-PECD depends not only on the transient bound resonances – as evidenced in Fig. 2(a-c) and Fig. 3(b-d) – but also on the dichroism encoded by ionization by circularly polarized light. In this respect, we note that a photoexcitation electron circular dichroism (PXCD) configuration[44] in which molecules are photoexcited by a circularly polarized pump pulse and subsequently ionized with a linearly polarized probe would reduce the degrees of freedom to only the transient bound resonances.

The peculiarity of the electron dynamics uncovered in this work yet underlies another very important consequence on the molecular response of photoexcited chiral systems: the coherent superposition of excited states induced by the pump allows to selectively filter - within few femtoseconds - specific molecular orientations through enantio-sensitive photoionization[13]. An electron moving within the coherent state superposition creates an electronic current $\boldsymbol{J}(\widehat{\boldsymbol{R}}, \boldsymbol{r}, t) = \frac{\hbar}{m}\Im[\Phi^*(\widehat{\boldsymbol{R}}, \boldsymbol{r}, t)\boldsymbol{\nabla}\Phi(\widehat{\boldsymbol{R}}, \boldsymbol{r}, t)]$ which reduces to

$$\boldsymbol{J}(\widehat{\boldsymbol{R}}, \boldsymbol{r}, t) = \frac{\hbar}{m} A_{3d}(\widehat{\boldsymbol{R}}) A_{4p}(\widehat{\boldsymbol{R}}) [\Psi_{4p}(\boldsymbol{r})\boldsymbol{\nabla}\Psi_{3d}(\boldsymbol{r}) - \Psi_{3d}(\boldsymbol{r})\boldsymbol{\nabla}\Psi_{4p}(\boldsymbol{r})] \sin\left[\frac{(E_{4p} - E_{3d})t}{\hbar}\right] \quad (4)$$

when expanding $\Phi$ on the (real) 3d and 4p bound eigenstates. The chirality of the molecule induces a curl in the generated electron current, whose rotation direction reverses periodically. Rotating electron currents are known to influence the ionization probability by circularly polarized light: the propensity rules[45] establish that 1-photon ionization is enhanced when the electrons rotate in the same direction as the electric field. Therefore, the molecules oriented such that their electronic current rotates in the same plane and direction as the ionizing laser pulse are preferentially ionized, see Fig. 4(d). Consequently, the produced molecular cations are selectively oriented along the probe polarization rotation axis, corresponding to the light propagation axis $\hat{\boldsymbol{z}}$.

To quantify the degree of orientation of the photoionized molecules, we select an unitary vector $\hat{\boldsymbol{e}}_{mol}$ fixed to the internal C-C bond of the ML cation, as illustrated in the inset of Fig.

4(e) (see also Fig. S10) and calculate its averaged value over the probe-filtered molecular orientations in the laboratory frame as[13]

$$\langle \hat{e}_{lab} \rangle_{\hat{R}}^{(\pm 1)}(\epsilon, t) = \frac{\int d\hat{R}\, W^{(\pm 1)}(\hat{R}, \epsilon, t) \hat{e}_{lab}(\hat{R})}{b_{0_{incoh}}^{(\pm 1)}(\epsilon)} \quad (5)$$

where $\hat{e}_{lab}(\hat{R})$ is the $\hat{e}_{mol}$ vector passively rotated in the laboratory frame and $W^{(\pm 1)}(\hat{R}, \epsilon, t)$ is the ionization rate associated with photoelectrons of energy $\epsilon$. Importantly, the averaged orientation of the cations depends on $\epsilon$ because the $\epsilon$-dependence of the underlying photoionization yield is not the same for all orientations $\hat{R}$. The x and y components of $\langle \hat{e}_{lab} \rangle_{\hat{R}}^{(\pm 1)}(\epsilon, t)$ are found to be zero and only the z-component survives the averaging[13], leading to

$$\langle \cos\theta_{ion} \rangle_{\hat{R}}^{(\pm 1)}(\epsilon, t) = \langle \cos\theta_{ion} \rangle_{cross}^{(\pm 1)}(\epsilon) \sin[(E_{4p} - E_{3d})t/\hbar] \quad (6)$$

where $\theta_{ion}$ is the angle between the internal C-C bond and the probe propagation $\hat{z}$ axis (see inset of Fig. 4(e)). $\langle \cos\theta_{ion} \rangle_{cross}^{(\pm 1)}(\epsilon)$ involves chiral-sensitive products of 3d and 4p excitation and ionization amplitudes. The temporal evolution of $\langle \cos\theta_{ion} \rangle_{\hat{R}}^{(+1)}$ is illustrated in Fig. 4(e) for $\epsilon = 250$ meV. When $\langle \cos\theta_{ion} \rangle_{\hat{R}}^{(+1)}(\epsilon, t) < 0$, the $CO_2CH_3$ moiety of the ML cations preferentially points forward with respect to $\hat{z}$ while it rather points backward when $\langle \cos\theta_{ion} \rangle_{\hat{R}}^{(+1)}(\epsilon, t) > 0$. Such asymmetry could be detected by resolving the direction of fragmentation of molecular cations. Indeed, the relative numbers of molecules pointing forward and backward at time $t$, $N_+^{(\pm 1)}(\epsilon, t)$ and $N_-^{(\pm 1)}(\epsilon, t)$, respectively, can be linked to $\langle \cos\theta_{ion} \rangle_{\hat{R}}^{(\pm 1)}(\epsilon, t)$ (see Methods). This ultrafast filtering of molecular orientation has consequences of paramount importance on subsequent reactive dynamics of ML cations. The prompt photoionization dictates the subsequent dissociation along the selected molecular orientation and a forward/backward fragment asymmetry (FBFA) naturally arises, which we define as

$$FBFA^{(\pm 1)}(\epsilon, t) = 2\frac{N_+^{(\pm 1)}(\epsilon,t) - N_-^{(\pm 1)}(\epsilon,t)}{N_+^{(\pm 1)}(\epsilon,t) + N_-^{(\pm 1)}(\epsilon,t)}. \qquad (7)$$

The FBFA is illustrated in Fig. 4(f) for $h = +1$ and $\epsilon = 250$ meV, reaching absolute values of ~30% while its temporal evolution is dictated by the behavior of the underlying electron current $J(\widehat{R}, r, t)$. Similarly, to the MP-PECD, the FBFA switches sign for $h = -1$ or when the other enantiomeric form of ML molecules is considered. Since the FBFA is created by the electron current, it vanishes in the case of incoherent population of excited states (Fig. 4(c)). This shows that the chiral electronic coherence directly observed in our experiment via TR-PECD is crucial to achieve control over enantio-selective dynamics of the nuclei.

To conclude, the potential of time-resolved PECD to probe transient chirality had so far only been demonstrated experimentally for nuclear dynamics, internal conversion and photoionization delays in chiral molecules [8–11,46]. We have taken an important step forward by resolving the coherent chiral electronic dynamics of a chiral molecule in the first instants following prompt excitation by an achiral few-fs UV pulse. The results demonstrate that TR-PECD can provide insights on the role of the primary electron dynamics in the light-induced chiral response of complex molecular systems such as chiral biomolecules and organometallic complexes. Offering a route to investigate the fundamental origin of chiral recognition that is ubiquitous in biological phenomena[47], the possibility to control the photoelectron emission direction in the laboratory frame also offers the potential to engineer petahertz switching devices based on chiral interactions. Finally, we demonstrated that beyond its impact on the chiroptical properties of the system, the chiral currents generated in our experiment can be exploited for enantio-sensitive charge-directed reactivity leading to oriented fragmentation. Steering the outcome of these photophysical and photochemical properties provides an important added-value to chirality at the molecular scale.

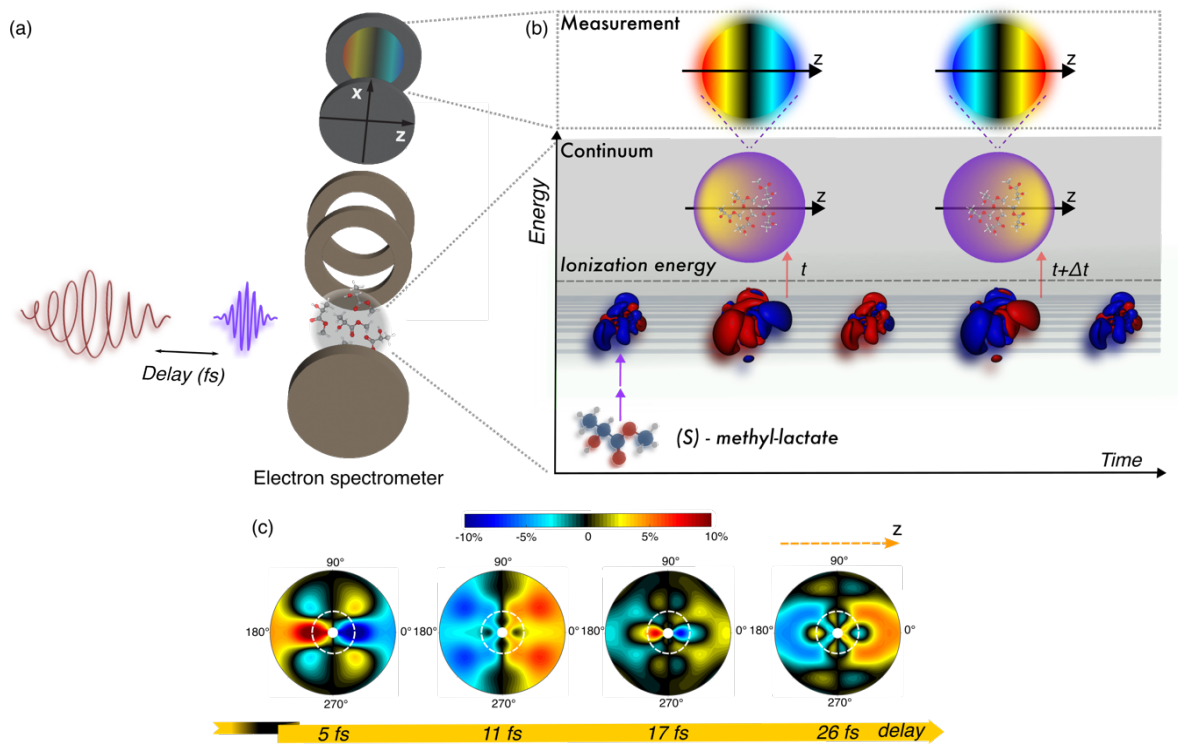

Fig. 1: **Light-induced chiral dynamics of methyl-lactate.** (a) A few-femtosecond linearly polarized UV pulse excites an ensemble of randomly oriented chiral molecules, creating an electronic wavepacket of Rydberg states via 2-photon absorption. The dynamics is probed via 1-photon ionization by a time-delayed circularly polarized NIR pulse. The probing step leads to the ejection of photoelectrons along the light propagation axis defined along the *z* direction and the resulting angular distribution is recorded by a velocity map imaging spectrometer. (b) The red and blue structure shows the temporal evolution of the coherent electron density in the excited neutral molecule: the chiral evolution of the photoexcited Rydberg wavepacket leads to a reversal of the 3D photoelectron angular distribution at two distinct time delays *t* and *t+Δt*, captured by the measurements. (c) For each time delay, an image is recorded for both left and right circular polarization of the probe pulse. The differential image PECD($\epsilon,\theta,t$) defined in the main text is shown for time delays of 5, 11,17 and 26 fs for photoelectrons with kinetic energies from 25 to 300 meV along the radial coordinate. The white circles identify the photoelectrons below 100 meV which experience an ultrafast reversal of their emission direction in the laboratory frame.

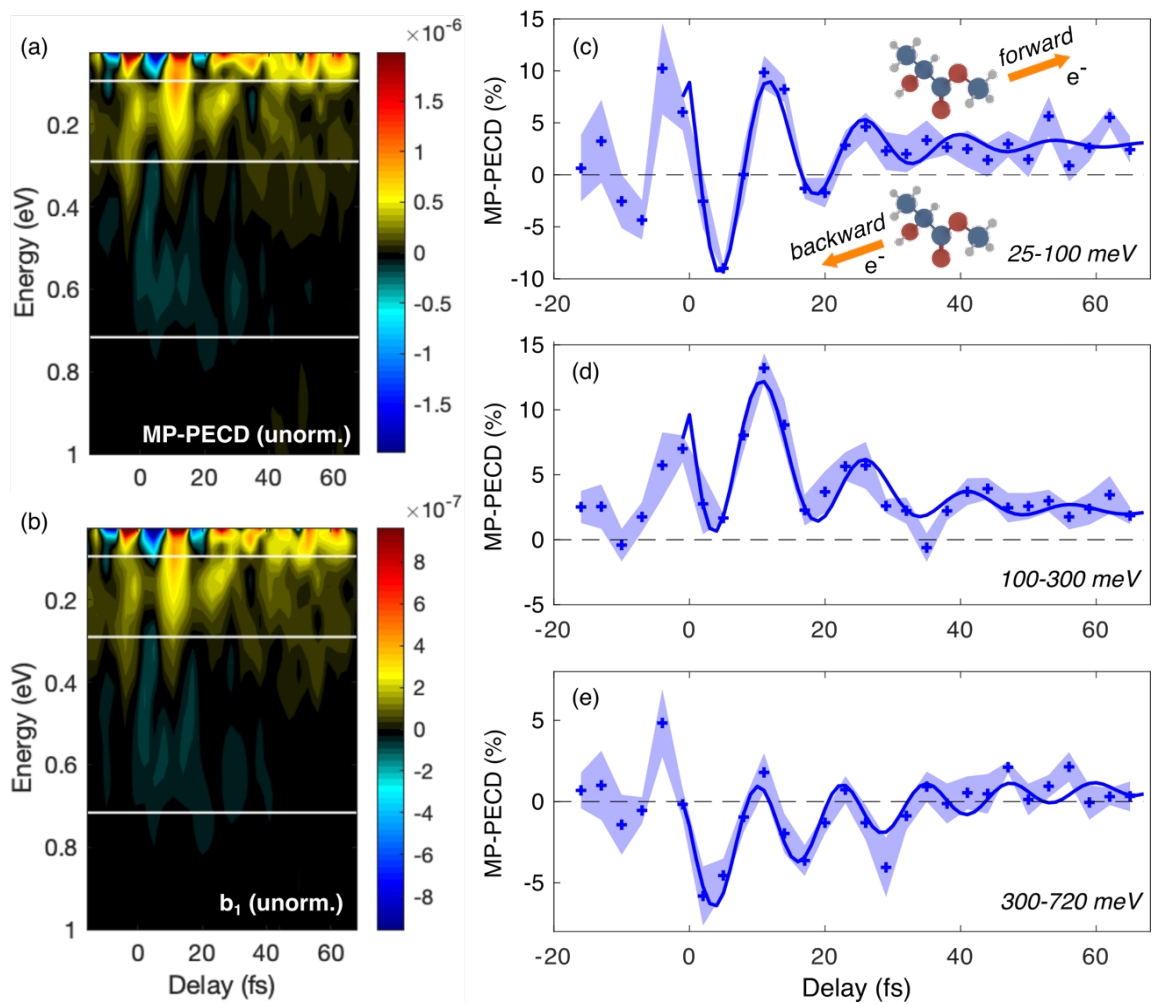

Fig. 2: **Energy-resolved analysis.** Temporal evolution of the unnormalized MP-PECD in (*S*)-methyl-lactate (a) and corresponding $b_1$ coefficient (b). The white lines identifiy three different kinetic energy ranges of photoelectrons: 25-100 meV (c), 100-300 meV (d) and 300-720 meV (e). The standard error of the mean over 5 measurements is shown by the shaded areas. The solid blue lines show the fit of the oscillations from *t* = 0 fs (see the corresponding Fourier analysis in Fig. 3c,e). The change of sign in (c) identifies a reversal of the photoelectron emission direction in the laboratory frame.

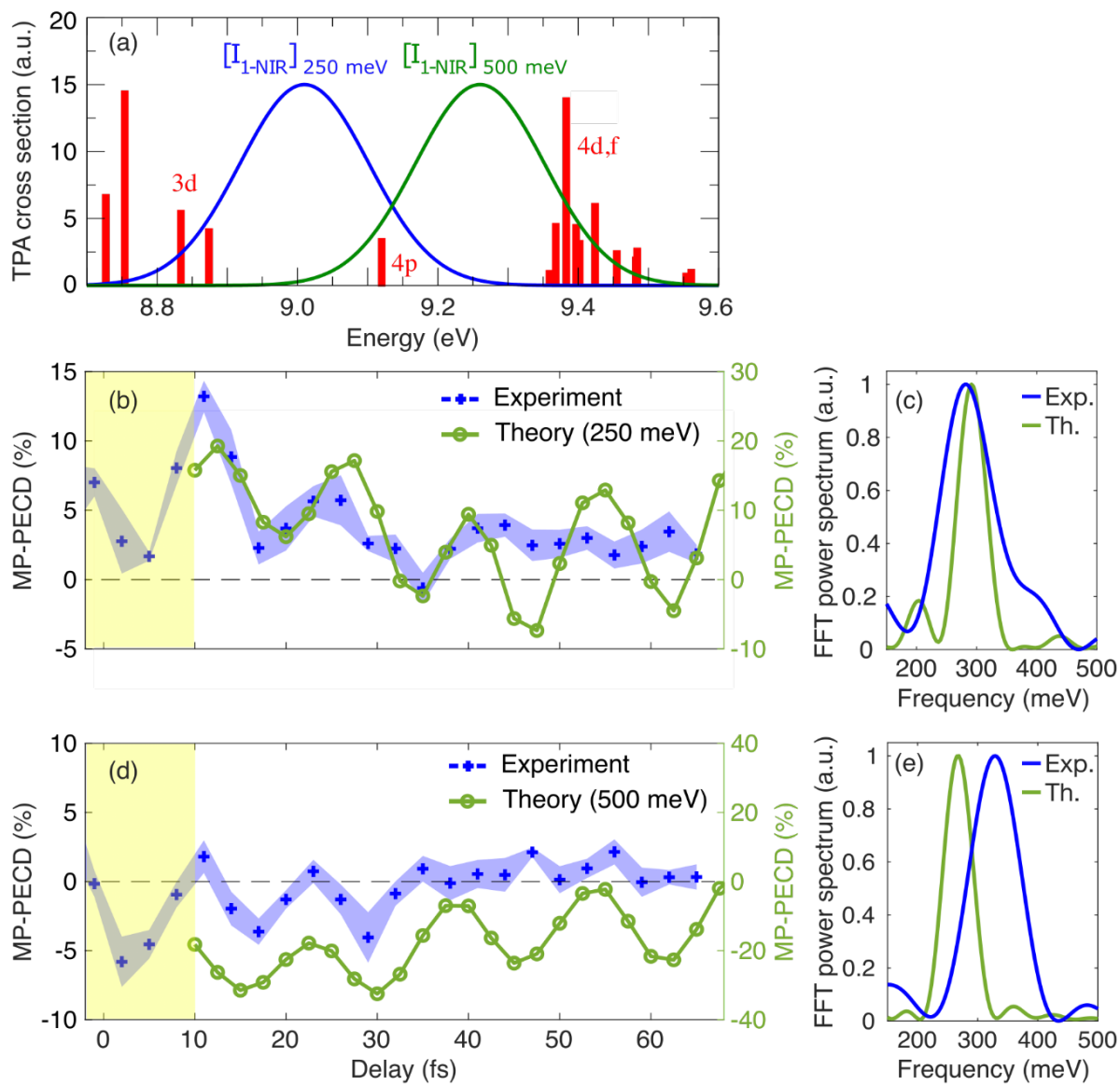

Fig. 3: **Modelling of the experiment.** (a) Two-photon absorption (TPA) cross sections for the excited states stemming from almost pure HOMO excitation. The cross sections have been convoluted with the UV-pump intensity squared. The blue and green curves correspond to the spectral probe intensity, down-shifted in energy to elicit the transient Rydberg states leading to photoelectrons with energies $\epsilon = 250$ and $\epsilon = 500$ meV through ionization by one photon centered at frequency $\omega = 1.75$ eV. (b) Calculated MP-PECD for photoelectrons with $\epsilon = 250$ meV (green) compared to the experiment (blue). The calculations start at $t = 10$ fs corresponding to the end of the pump-probe overlap region (yellow area). (c) Corresponding power spectra from a Fourier analysis. The frequency axis is displayed for beatings of excited states with an energy spacing between 150 meV (27.6 fs period) and 500 meV (8.3 fs). The main peak from the computed MP-PECD evolution is at 291 meV (14.2 fs). The power spectrum of the experimental data shows a peak frequency at 280 meV (14.8 fs) and was evaluated up to $t = 35$ fs where the oscillations are damped. (d) Calculated MP-PECD for photoelectrons with $\epsilon = 500$ meV (green) compared to the experiment (blue). (e) Corresponding power spectra with a central component at 269 meV (15.4 fs) for the computed curve. The power spectrum of the experimental data is shown with a central frequency at ~329 meV (12.6 fs).

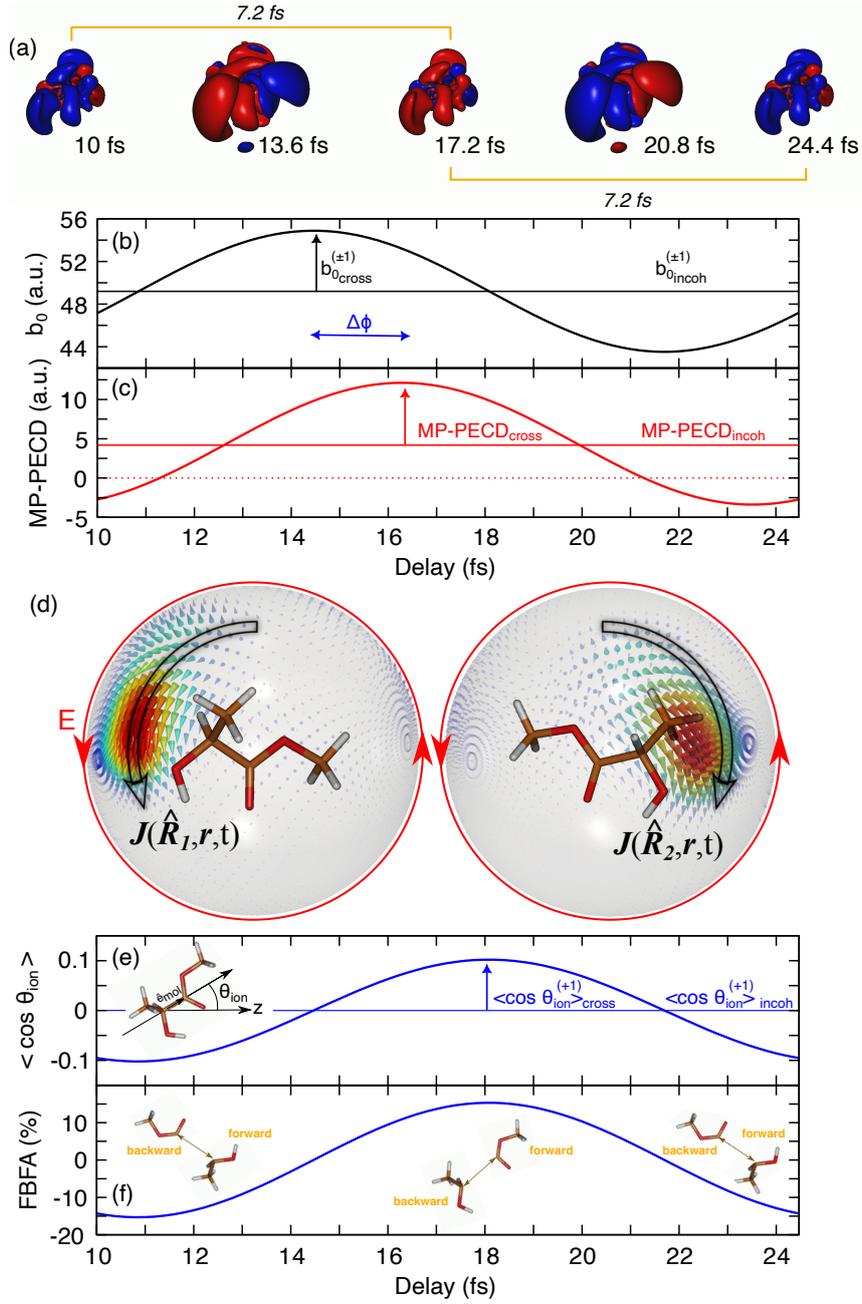

Fig. 4: **Electron-driven dynamics in the case of quantum beating monitored by 3d and 4p Rydberg states**. (a) Temporal evolution of the coherent part of the electron density over one period of the quantum beating between the 3d and 4p states (see equation (1) of the text). (b) Photoelectron yield as a function of the pump-probe delay for $\epsilon = 250$ meV. The Rydberg quantum beating leads to an oscillatory behavior of the yield which is in phase with the variation of the electron density shown in (a), as expected from equations (1) and (2). (c) MP-PECD as a function of the pump-probe delay for $\epsilon = 250$ meV. The dichroism is delayed by $\Delta\phi = 0.79$ rad (1.8 fs) with respect to the variation of the electron density because of the interferences between the continuum partial wave amplitudes (see equation (3)). (d) Snapshots of the electronic current induced by the pump pulse, on a Rydberg sphere of 10 a.u. radius surrounding the molecule for two distinct orientations $\hat{R}_i$. The current is defined in the molecular frame (equation (4)) and ionization by the probe pulse is enhanced for the molecular orientation where (i) the current co-rotates with the circularly polarized probe field (red arrow) and (ii) the rotation axis of the current aligns with the light propagation vector. This happens here only for $\hat{R}_1$. (e) Active orientation of the produced cations along the light propagation axis $\hat{z}$ as a function of time for $\epsilon = 250$ meV. This orientation is defined as the mean value of $cos\,\theta_{ion}$, where $\theta_{ion}$ is the angle between the internal C-C bond of the ML cation and $\hat{z}$, as shown in the inset (see equation (6)). (f) The resulting forward/backward fragment asymmetry along $\hat{z}$ in the reactive fragmentation of ML cations, following probe-induced ionization of the transient 3d-4p electron wavepacket leading to photoelectrons with energy $\epsilon = 250$ meV (see equation (7)). The insets illustrate the preferential directions of emission of $CO_2CH_3$ and $CH_3CHOH^+$ fragments.

**METHODS**

**Experimental setup**

The experiments were carried out with a 1 kHz titanium:sapphire laser (FemtoPower, Spectra-Physics), delivering 25-fs, 12-mJ pulses at 800 nm. 5.6 mJ were used for spectral broadening in a 2.3-m long hollow-core fiber (few-cycle inc.) filled with a pressure gradient of helium gas. The fiber setup seeds an all-vacuum Mach-Zehnder-like interferometer with 5-fs near-infrared (NIR) pulses. One arm is used for the generation of the UV-pump pulse via third-harmonic generation in a laser-machined glass cell filled with 7.2 bar of neon gas. A pair of silicon superpolished substrates (Gooch & Housego) is used at Brewster angle to attenuate the residual part of the NIR driving field by 3 orders of magnitude while reflecting ~16% of the UV radiation (50 nJ). In the second arm of the interferometer, the remaining part of the NIR beam is focused to the experimental region by a toroidal mirror ($f$ = -900 mm) followed by a motorized zero-order quarter-waveplate (B. Halle) in order to control the helicity of the circularly polarized probe pulses (16 µJ), with an intensity of $5\times10^{12}$ Wcm$^{-2}$. The instrument response function of 2.90 ± 0.06 fs is obtained by a global fit of the non-resonant (gaussian) dynamics of all the ion masses acquired simultaneously with the photoelectron spectra (see SI). Liquid (*S*)-methyl-lactate (97% enantiomeric excess, Sigma-Aldrich) was evaporated and transported by diffusion to a velocity map imaging spectrometer to measure the photoelectron angular distribution as a function of the pump-probe time delay. To avoid condensation of the sample along the transportation line within the molecular source, a temperature gradient from 85°C to 95°C was applied.

**Analysis of the VMI images**

For each pump-probe delay $t$, the photoelectron angular distributions are collected with a velocity map imaging spectrometer (VMIS) for both left ($h = +1$) and right ($h = -1$) circular polarizations of the NIR probe pulse to yield $S^{(h)}(\epsilon, \theta, t)$, where $\epsilon$ is the kinetic energy of the photoelectron and $\theta$ its emission angle with respect to the light propagation axis. The differential PECD image is then defined as the normalized difference $PECD(\epsilon, \theta, t) = 2\frac{S^{(+1)}(\epsilon,\theta,t) - S^{(-1)}(\epsilon,\theta,t)}{S^{(+1)}(\epsilon,\theta,t) + S^{(-1)}(\epsilon,\theta,t)}$, subsequently fitted using a pBasex inversion algorithm[8]. Its evolution is monitored as a function of the pump-probe delay $t$ in Fig. 1c. $S^{(h)}(\epsilon, \theta, t) = \sum_{n=0}^{2N} b_n^{(h)}(\epsilon, t) P_n(\cos\theta)$ where $P_n(\cos\theta)$ are Legendre polynomials and $N = 3$ is the total number of photons absorbed to reach the continuum from the ground state: the pump-induced excitation involves two photons while ionization consists in the absorption of one NIR probe photon. $b_0^{(h)}(\epsilon, t)$ corresponds to the total (angle-integrated) photoionization cross section. In the case of a sample of randomly oriented achiral molecules, the PAD is symmetric with respect to the light propagation axis so that the $S^{(h)}(\epsilon, \theta, t)$ expansion is restricted to even $n$'s. For randomly oriented chiral molecules, the asymmetric contribution to the photoelectron yield emerges from the additional $b_n^{(h)}$ amplitude coefficients with odd $n$. Besides $PECD(\epsilon, \theta, t)$, it is convenient to introduce an angularly-integrated quantity to characterize the whole chiroptical response at fixed kinetic energy. Defining it as the difference of electrons emitted in the forward and backward hemispheres for $h = +1$, normalized to the average number of electrons collected in one hemisphere, we obtain the so-called multiphotonic (MP)-PECD[28], $MP - PECD(\epsilon, t) = 2\beta_1^{(+1)}(\epsilon, t) - \frac{1}{2}\beta_3^{(+1)}(\epsilon, t) + \frac{1}{4}\beta_5^{(+1)}(\epsilon, t)$ where $\beta_n^{(+1)}(\epsilon, t) = \frac{b_n^{(+1)}(\epsilon,t)}{b_0^{(+1)}(\epsilon,t)}$. The time and energy resolved amplitude coefficients $\beta_n^{(+1)}(\epsilon, t)$, together with the resulting unnormalized $MP-PECD(\epsilon, t)$, are shown in Fig. S2 of the SI. Note that $\beta_5^{(+1)}(\epsilon, t)$ is not included due to its negligible contribution to the total signal. The validity of the analysis protocol despite the lack of cylindrical symmetry induced by the anisotropy of excitation of the linearly polarized UV-pump pulse has been demonstrated in [8,9].

**Computation of TR-PECD**

At time $t$ after the pump pulse vanishes, the electron wavepacket formed in a ML molecule whose orientation in the laboratory frame is characterized by $\widehat{R}$ reads

$$\Phi(\widehat{R}, r, t) = \sum_i \mathcal{A}_i(\widehat{R}) \Psi_i(r) \exp(-iE_i t/\hbar),$$

where $\Psi_i(r)$ are excited states with energies $E_i$ and two-photon absorption amplitudes from the ground state $\mathcal{A}_i(\widehat{R})$. These states, energies and transition amplitudes, have been obtained by large-scale TDDFT[29] calculations, detailed in Section 2.1 of the SI. In the spectral region spanned by the pump pulse, most of the excited states have a Rydberg character and stem from the excitation of the ML HOMO (see Fig. S5 of the SI).

The absorption of one NIR photon of the probe pulse leads to the ejection of a photoelectron with wavevector $\widehat{k}'$ in the molecular frame. The associated ionization dipole is

$$d_{k'}^{h,mol}(\widehat{R}, t) = \sum_i \mathcal{A}_i(\widehat{R}) \sqrt{I_{1-NIR}(\omega_i)} <\Psi_{k'}^{(-)}|\widehat{e_h} \cdot r|\Psi_i> \exp(-iE_i t/\hbar)$$

where $I_{1-NIR}(\omega_i)$ is the spectral intensity of the probe pulse at frequency $\omega_i = k'^2/2 + I_p - E_i$, with $I_p$ the ML ionization potential, $\widehat{e_h}$ is the circular polarization of the probe pulse ($h = \pm 1$) and $\Psi_{k'}^{(-)}$ is the ingoing scattering state associated with the electron ejected in the continuum. Neither the scattering state nor the excited states explicitly depend on $t$ since the calculations are made assuming that the nuclei remain frozen at their equilibrium locations at all $t$ (see Section 2.3.1 of the SI). $\Psi_{k'}^{(-)}(r)$ is obtained in the framework of the X$\alpha$ approximation for the exchange electron interaction[30,31], detailed in Section 2.2.2 of the SI.

Rotating the ionization dipole into the laboratory frame allows us to define the orientation-averaged differential ionization cross section as

$$\frac{d\bar{\sigma}^{(h)}}{d\Omega_k}(k,\theta,\varphi,t) \propto \int d\widehat{R}|d_k^{h,lab}(\widehat{R},t)|^2$$

where $k = k'$ and $(\theta, \varphi)$ are the spherical angles characterizing the direction $\widehat{k}$ of electron ejection in the laboratory frame -- $\theta$ is defined with respect to the pulse propagation direction $\widehat{z}$. While the cross section can be put in the closed form

$$\frac{d\bar{\sigma}^{(h)}}{d\Omega_k}(k,\theta,\varphi,t) = \sum_{s=0}^{6}\sum_{i=-2}^{2} b_{s,2i}^{(h)}(k,t)Y_s^{2i}(\theta,\varphi),$$

where $Y_s^{2i}(\theta,\varphi)$ are spherical harmonics, we show in section 2.2.2 of the SI that the MP-PECD is

$$MP - PECD(\epsilon, t) = \frac{1}{b_{0,0}^{(+1)}(\epsilon,t)}\left(2\sqrt{3}b_{1,0}^{(+1)}(\epsilon,t) - \frac{\sqrt{7}}{2}b_{3,0}^{(+1)}(\epsilon,t) + \frac{\sqrt{11}}{4}b_{5,0}^{(+1)}(\epsilon,t)\right).$$

where $\epsilon = \frac{\hbar^2 k^2}{2m}$, with $m$ the electron mass, is the photoelectron kinetic energy. The $b_{s,2i}^{(h)}$ coefficients basically depend on partial-wave ionization amplitudes weighted by the primary excitation factors, as shown in Section 2.2.2 of the SI.

**Ab initio molecular dynamics**

Nuclear dynamics are described in a classical framework using the Newtonian equation of motion

$$M_i\ddot{R}_i(t) = -\nabla_i V_e(\{R_j(t)\})$$

where $R_j(t)$ is the coordinate vector of nucleus $j$ with mass $M_j$, and $V_e(\{R_j(t)\})$ is the potential arising from the electrons within the ground state of the ML cation. The initial coordinates $\{R_j(t=0)\}$ and momenta $\{P_j(t=0) = M_j \dot{R}_j(t=0)\}$ of the nuclei are randomly taken from a Wigner distribution of the harmonic vibrational ground state of neutral ML in its ground electronic state. The Wigner distribution is discretized in terms of 250 sets $\{\{R_j(t=0)\},\{P_j(t=0)\}\}_{k=1,..,250}$ which are the starting points of 250 non-interacting trajectories $\{\{R_j(t)\},\{P_j(t)\}\}_{k=1,..,250}$. The trajectories are propagated using the Newton-X package[48,49] and the electronic potential $V_e(\{R_j(t)\})$ is evaluated 'on the fly' using DFT at the CAM-B3LYP/6-311++G(d,p) level of theory[50–52]. Early stage ($t \leq 60$ fs) and subsequent nuclear dynamics are discussed and illustrated in Sections 2.3 and 2.4.2 of the SI, respectively. Note that the sudden electron excitation and ionization processes mainly occur in classically allowed regions of the potential energy surfaces. The nuclei stay in these regions for short pump-probe delays of a few tens of femtoseconds, which is the timescale of interest in our work. The classical treatment of nuclear dynamics is thus appropriate.

**Probe-induced active orientation of the sample and enantio-selective reactive dynamics**

The ionization rate $W^{(\pm 1)}(\widehat{R}, \epsilon, t)$ involved in the averaged value of the probe-filtered molecular orientation in the laboratory frame, $\langle \hat{e}_{lab} \rangle_{\widehat{R}}^{(\pm 1)}(\epsilon, t)$ in equation (5), is

$$W^{(\pm 1)}(\widehat{R}, \epsilon, t) \propto \int |d_k^{\pm 1, lab}(\widehat{R}, t)|^2 d\widehat{k}.$$

Its expression, involving primary excitation and ionization amplitudes, is detailed in Section 2.4 of the SI.

The z-component of $\langle \hat{e}_{lab} \rangle_{\widehat{R}}^{(\pm 1)}(\epsilon, t)$, $\langle \cos\theta_{ion} \rangle_{\widehat{R}}^{(\pm 1)}(\epsilon, t)$, is given in equation (6). Its relation with the number of molecules pointing forward and backward with respect to the light propagation axis, $N_+^{(\pm 1)}(\epsilon, t)$ and $N_-^{(\pm 1)}(\epsilon, t)$, respectively, is derived from the simple orientation model introduced in[13]: the distribution of oriented molecules is described by the wavefunction $\chi(\theta_{ion}, \epsilon, t) = a_0(\epsilon, t)Y_0^0(\theta_{ion}) + a_0(\epsilon, t)Y_1^0(\theta_{ion})$ where $Y_0^0(\theta_{ion})$ and $Y_1^0(\theta_{ion})$ are the usual spherical harmonics and $a_0^2(\epsilon, t) + a_1^2(\epsilon, t) = 1$. Note that higher $(l, m)$ orders are not necessary to describe basically a forward/backward asymmetry. $\langle\cos\theta_{ion}\rangle_{\widehat{R}}^{(\pm 1)}(\epsilon, t)$ can be evaluated using the $\chi(\theta_{ion}, \epsilon, t)$ expansion as $\langle\theta_{ion}\rangle_{\widehat{R}}^{(\pm 1)}(\epsilon, t) = \int_0^{2\pi} d\varphi \int_0^{\pi} d\theta_{ion} |\chi(\theta_{ion}, \epsilon, t)|^2 \cos(\theta_{ion})$, yielding $\langle\cos\theta_{ion}\rangle_{\widehat{R}}^{(\pm 1)}(\epsilon, t) = \frac{2}{\sqrt{3}} a_0(\epsilon, t) a_1(\epsilon, t)$.

The relative number of molecules pointing forward simply corresponds to the square modulus of $\chi(\theta_{ion}, \epsilon, t)$ integrated over the forward hemisphere where $\theta_{ion} \in \left[0, \frac{\pi}{2}\right]$:

$$N_+^{(\pm 1)}(\epsilon, t) = \int_0^{2\pi} d\varphi \int_0^{\pi/2} d\theta_{ion} |\chi(\theta_{ion}, \epsilon, t)|^2 = \frac{1}{2} + \frac{\sqrt{3}}{2} a_0(\epsilon, t) a_1(\epsilon, t) = \frac{1}{2} + \frac{3}{4}\langle\cos\theta_{ion}\rangle_{\widehat{R}}^{(\pm 1)}(\epsilon, t).$$

Obviously $N_-^{(\pm 1)}(\epsilon, t) = 1 - N_+^{(\pm 1)}(\epsilon, t)$. The explicit expression of $\langle\cos\theta_{ion}\rangle_{\widehat{R}}^{(\pm 1)}(\epsilon, t)$ is detailed in Section 2.4 of the SI. Once $N_+^{(\pm 1)}(\epsilon, t)$ is known, the FBFA defined in equation (7) directly follows.

**ACKNOWLEDGEMENTS**

We thank F. Remacle for fruitful discussions and her constructive feedback and K. Pikull for the excellent technical support.

**AUTHOR CONTRIBUTIONS**

V.W., E.B., V.B., Y.M., B.P. and F.C. conceived the experiment. V.W., E.B., E.P.M., L.C., S.R., K.S. and A.T. performed the experiments. V.W., E.B. and Y.M., carried out the data analysis. M.C.H., N.B.A. and B.P. calculated the molecular and electronic properties of methyl-lactate. B.P. performed the MP-PECD and molecular orientation calculations. M.C.H. performed the classical trajectory simulations. A.F.O., D.A. and O.S. identified and developed the concept of enantio-sensitive molecular orientation. V.W., Y.M., B.P. and F.C. drafted the manuscript. All authors contributed to the discussion of the results and the editing of the manuscript.

**COMPETING INTERESTS**

The authors declare no competing interests.

**FUNDING**

We acknowledge financial support from the European Research Council under the ERC-2014-StG STARLIGHT (grant no. 637756), European Union's Horizon 2020 research and innovation program No. 682978 – EXCITERS, the German Research Foundation (DFG)—SFB-925—project 170620586 and the Cluster of Excellence Advanced Imaging of Matter (AIM).

**DATA AVAILABILITY**

The data that support the findings of this study are available from the corresponding author upon reasonable request.

**CODE AVAILABILITY**

The code used for the simulations contained in this study is available from the corresponding authors upon reasonable request.